\documentclass[12pt,preprint]{aastex}

\usepackage{amsmath}	
\usepackage{amssymb}	

\newcommand{\be}{\begin{equation}}
\newcommand{\ee}{\end{equation}}
\newcommand{\bea}{\begin{eqnarray}}
\newcommand{\eea}{\end{eqnarray}}
\newcommand{\bal}{\begin{aligned}}
\newcommand{\eal}{\end{aligned}}

\begin{document}

\title{``Firewall'' Phenomenology with Astrophysical Neutrinos}

\author{Niayesh Afshordi and Yasaman K. Yazdi}
\affil{Perimeter Institute for Theoretical Physics, 31 Caroline St. N., Waterloo ON, N2L 2Y5, Canada}

\affil{Department of Physics and Astronomy, University of Waterloo, Waterloo ON, N2L 3G1, Canada}

\email{nafshordi@pitp.ca}
\email{yyazdi@pitp.ca}

\begin{abstract}
One of the most fundamental features of a black hole in general relativity is its event horizon: a boundary from which nothing can escape. There has been a recent surge of interest in the nature of these event horizons and their local neighbourhoods. In an attempt to resolve black hole information paradox(es), and more generally, to better understand the path towards quantum gravity, ``firewalls" have been proposed as an alternative to black hole  event horizons. In this paper, we explore the phenomenological implications of black holes possessing a surface or ``firewall'', and predict a potentially detectable signature of these firewalls in the form of a high energy astrophysical neutrino flux. We compute the spectrum of this neutrino flux in different models and show that it is a possible candidate for the source of the PeV neutrinos recently detected by IceCube. This opens up a new area of research, bridging the non-perturbative physics of quantum gravity with the observational black hole and high energy astrophysics. 
\end{abstract}




\section{Introduction}\label{intro}
One of the most celebrated predictions of General Relativity has been the possibility of forming black holes: spacetime singularities that are surrounded by event horizons. These event horizons are boundaries in the spacetime from which nothing can escape. In contrast to black holes, other astrophysical compact objects such as neutron stars, possess a physical surface that shows visible signs of radiation to the outside universe.  In this work we explore the possibility that black holes, too, may not be so invisible after all.

There has been much recent interest in studying the nature of black hole event horizons and their local neighbourhoods, in an attempt to resolve black hole information paradox(es), and more generally to better understand the path towards quantum gravity \citep[e.g.,][]{firewall1, y1, y2, y3, y4, y5, y6, y7, y8, y9, y10, y11, y12, y13, y14, y15, y16, y17, y18, y19, y20, y21, y22, y23, y24, y25, y26, y27, y28, y29, y30, y31, y32,y33, y34, mehdi}. In particular, alternatives to classical black hole event horizons such as {\it firewalls} \citep{firewall1}, {\it fuzzballs}, \citep[e.g.,][]{y32}, or other exotic byproducts of quantum gravitational collapse	\citep[e.g.,][]{2014PhRvD..90l4033G,2015arXiv150405536D} have been discussed in the literature. It is generally believed that, in lieu of an event horizon, accretion onto (or scattering off) such exotic byproducts {\it may} have astrophysical observable consequences that distinguish them from classical black holes \citep{avery,y4,2015ApJ...805..179B}. 

We know that a quantum theory of gravity is non-local on the scale of the Planck length/time, if not longer, which means that a radical change in the structure of space-time within a Planck length of the horizon can affect the microscopic vicinity outside the event horizon. In this work, we assume that gravitational collapse leads to a ``firewall'' (a term we shall use as a placeholder for any exotic alternative to classical event horizons), that can show visible signs of emission to the outside universe. Such emission may be possible because an {\it accreting} firewall is not in global thermal equilibrium, as the timescale for global thermal equilibrium is generically much longer than the timescale for local radiative processes for macroscopic objects. This is analogous to meteorites hitting the Earth's atmosphere, where light is emitted from the meteor (within seconds), while the long timescale of heating up the entire earth and global thermal equilibrium are irrelevant.

We construct a phenomenological model to describe and constrain the nature of this radiation and its spectrum.  A key feature of our prediction for this radiation is that it will primarily consist of neutrinos. This is because, as in core-collapse supernovae, which nominally share similar radiative properties to firewalls (e.g. surface density and composition), neutrino propagation has a much smaller optical depth and therefore transports energy out more efficiently. 

An interesting implication is that accretion onto black holes/firewalls could be a possible source for the recently detected high energy (PeV) neutrinos by IceCube \citep{icecube3}. The source of this detection is as yet unknown, and several candidate sources have already been ruled out or strongly constrained. For example, hadronuclear $pp$ processes (the mechanism of neutrino production in intergalactic shocks, starburst galaxies, and some active galactic nuclei (AGN) models) require power laws harder ($p\leq 2.1-2.2$)  than those favored by IceCube's detection \citep{murase}. Inner jets of blazars may be able to account for IceCube's PeV signal, but they have difficulty explaining the sub-PeV neutrino events \citep{murase2}. Further constraints on the candidate sources are set by arguments that no extragalactic object that we know of could function as the origin of both the PeV neutrinos and the ultra high energy cosmic rays \citep{yoshida}.

We start by outlining the main features and assumptions in our model in Sec. \ref{sec:model}. We then summarize our estimate of the total flux emitted by the black holes (Sec. \ref{sec:flux}) followed by modelling their spectra (Sec. \ref{sec:spectrum}). While we consider both blackbody and power law spectra, we subsequently focus on the power law spectrum in Sec. \ref{sec:results} (which are  more interesting at higher energies), comparing our results to observations and other models. Finally, Sec. \ref{sec:conclusions} concludes the paper. 

\section{Modelling Accretion onto a firewall}\label{sec:model}
The black holes in our model are accretion powered, have Keplerian flows, and have reached approximate steady state.

We assume that the accreting black holes must show evidence of surface radiation. \citet{avery} considered the possibility of photon emission from the surface of Sagittarius $A^*$. As their predicted photon flux exceeds the Sag. $A^*$ flux in millimetre and infrared observations,  they conclude that there must be an event horizon. In this work, we argue that at the energy scales involved, the neutrino cross-section is much smaller than the photon cross-section, hence (similar to core-collapse supernovae) most of the energy dissipated in an optically thin region will be carried out by the neutrinos. Based on this assumption, we go on to estimate the total flux of neutrinos from all the black holes (or firewalls) in the universe.

We estimate that a fraction ${\epsilon_g}$ (where $0<\epsilon_g\leq1$) of the gravitational binding energy of particles falling onto the surface of the black hole is radiated away (and reaches infinity), while the rest goes into growing the mass of the black hole. For a continuous accretion flow, as we are assuming here, this gives a total luminosity as measured at infinity of:


\be
L_{\infty}=\epsilon_g \dot{M} c^2.
\label{linf}
\ee

A fraction of the total luminosity, $L_{\rm \gamma}$, is converted by the accretion disc into electromagnetic radiation:
\be
L_{\rm \gamma}=\eta L_{\infty}.
\ee
The remainder of $L_{\infty}$, will be eventually radiated at the surface as neutrinos.  We call this surface luminosity as measured at infinity $L_{\nu}$:
\be
L_{\nu}\equiv L_{\infty}-L_{\gamma}=\epsilon_\nu \dot{M} c^2,
\label{surf}
\ee
where we have introduced a new efficiency $\epsilon_\nu\equiv \epsilon_g-\epsilon_g\eta$ for the surface radiation that we will be interested in for this work. In terms of the observed total integrated flux:\
\be
F_{\nu}= \epsilon_\nu \frac{\dot{M} c^2}{4\pi d_L^2},
\label{fsurf}
\ee
where $d_L$ is the luminosity distance. Now we want to sum over \eqref{fsurf} to find the total flux contribution from all the black holes in the observable universe. We treat the contributions from the stellar mass black holes and supermassive black holes separately. We estimate that $3.3\%$ \citep{cei} of the stellar mass budget goes into forming  stellar mass black holes, while  $M_{\rm BH}\sim 10~M_{\odot}$ and $M_{\rm SMBH}\sim 10^6 ~M_{\odot}$  are typical masses of stellar mass and supermassive black holes respectively. As a typical accretion rate we take $\dot{M} \sim 0.1~\dot{M}_{\rm Edd} = 0.1\times 16L_{\rm Edd} c^{-2}$, where $L_{\rm Edd}$ is the Eddington limit for the black hole mass.

\section{Total Flux}\label{sec:flux}

Let us make the following definitions of characteristic radii
\bea
&&r_h= \frac{GM}{c^2} \left(1+\sqrt{1-{a_*}^2}\right), \\
&&\alpha_1=1+(1-a_*^2)^{1/3}((1+a_*)^{1/3}+(1-a_*)^{1/3}), \\
&&\alpha_2=\sqrt{3 a_*^2+\alpha_1^2}, \\
&&r_{\rm ISCO}=\frac{GM}{c^2} \left[3+\alpha_2-\sqrt{(3-\alpha_1) (3+\alpha_1+2\alpha_2)}\right],\nonumber\\
\eea
where $r_h$ is the location of the event horizon, and $r_{\rm ISCO}$ is the radius of the innermost stable circular orbit. For simplicity, we assume that the black holes in our model are typically spinning with a dimensionless spin parameter of $a_*\sim 0.7$.

For the disc efficiency we use \cite{king}:
\be
\eta=\frac{1}{\epsilon_g}\sqrt{1-\frac{2 M}{3 r_{\rm ISCO}}}.
\ee

For example, for $\epsilon_g=0.5$, we will have efficiencies of $\eta=0.2$ and $\epsilon_\nu=0.4$ for both stellar mass and supermassive black holes. 

We primarily use models from \citet{SH2} and \citet{SH1}, for the evolution of the star formation rate (SFR) and the supermassive black hole accretion rate (BHAR) comoving densities, although our final results are not sensitive to our choice of these models. Both SFR and BHAR are fitted by a double exponential function:
\be
\dot{\rho}(z)=\dot{\rho}_0 \frac{b \ e^{a(z-z_m)}}{b-a+a \ e^{b(z-z_m)}},
\label{rho}
\ee
where the best fit parameters for the SFR are $a=3/5,\ b=14/15,\ z_m=5.4,\ \dot{\rho}_0=0.15~ M_{\odot} yr^{-1}  Mpc^{-3}$, and for the BHAR they are $a=5/4,\ b=3/2,\ z_m=4.8,\ \dot{\rho}_0=3\times 10^{-4}~M_{\odot} yr^{-1} Mpc^{-3}$.

The total flux then is

\be
\begin{split}
F^{\rm BH}_{\nu, tot}[z_{\rm max}]&= \frac{\epsilon_\nu c}{ H_\circ}\int_{0}^{z_{\rm max}} (0.033~ \dot{\rho}_{\rm SFR} c^2) \\
& \times \frac{1}{4\pi(1+z)^2\sqrt{\Omega_\Lambda+(1+z)^3\Omega_M}} \ dz, 
\label{ftotbhe}
\end{split}
\ee
\be
\begin{split}
F^{\rm SMBH}_{\nu, tot}[z_{\rm max}]&= \frac{\epsilon_\nu c}{H_\circ} \int_{0}^{z_{\rm max}} \dot{\rho}_{\rm BHAR}c^2 \\
&\times \frac{1}{4\pi(1+z)^2\sqrt{\Omega_\Lambda+(1+z)^3\Omega_M}} \ dz,
\label{ftotsmbhe}
\end{split}
\ee
where we use $H_\circ=73 km s^{-1}Mpc^{-1}$, $\Omega_\Lambda=0.7$, and $\Omega_M=0.3$.
We are working in comoving coordinates, so the volume elements in \eqref{ftotbhe} and \eqref{ftotsmbhe} are
\be
dV=\frac{c}{H_\circ}\frac{(1+z)^2 d_a^2}{\sqrt{\Omega_\Lambda+(1+z)^3\Omega_M}} \ dz,
\ee
where $d_A=\frac{c}{H_\circ(1+z)}\int_{0}^{z_{\rm max}}\frac{1}{\sqrt{\Omega_\Lambda+(1+z)^3\Omega_M}}\ dz$ and $d_L=(1+z)^2 d_A$.

For $z_{\rm max}=9$, near the redshift of reionization, Eq's \eqref{ftotbhe} and \eqref{ftotsmbhe} yield neutrino fluxes
\bea
F^{\rm BH}_{\nu, tot}[z<9] &\simeq& 3.8\times10^{-4} \ {\rm erg}\ {\rm s}^{-1} {\rm cm}^{-2} {\rm sr}^{-1}, \\
F^{\rm SMBH}_{\nu, tot}[z<9] &\simeq& 8.9\times 10^{-6} \ {\rm erg}\ {\rm s}^{-1} {\rm cm}^{-2} {\rm sr}^{-1}.
\eea

\section{Spectrum}\label{sec:spectrum}
To compare to observations, we need information about the spectrum of the neutrino flux. We consider two different models: a power law spectrum and a blackbody spectrum.
\subsection{Power Law with $p\sim2-3$}

For the power law model, the observed neutrino phase space density $f_{tot}[E_\nu]$ is
\be
f_{tot}[E_\nu]=\int^{z_{\rm max}} \left(\frac{\dot{\rho}(z) c^2 dV}{L_{\nu}(p)}\right)  \left(\frac{\pi r_A^2}{4\pi d_A^2}\right)  \left(\frac{E_\nu(1+z)}{E_\circ}\right)^{-p-2},
\ee
where the value of $E_\circ$ is fixed by the luminosity:

\be
L_{\nu}(p)=4\pi r_A^2\ \int_{E_\circ}^{E_P}\frac{ \ E_\nu^3}{\pi^2c^2\hbar^3} \left(\frac{E_\nu}{E_\circ}\right)^{-p-2}\ dE_\nu,
\ee
$r_A$ is the apparent radius of the black hole, and $E_P=1.96\times10^{16}$ erg is the Planck energy (even though the result is insensitive to this choice for $p>2$). Note that, given that the neutrino phase space density cannot exceed unity due to the Pauli exclusion principle, $E \sim E_\circ$ is the natural lower cut-off for a power-law spectrum. Details of how to obtain $r_A$ can be found in \cite{avery}: for $a_*\sim 0.7$ we have $r_A\sim5 GM c^{-2}$. 

The results depend on the precise choice of index $p$, which is one of our parameters. We will focus on $p\sim2-3$. The number flux is
\be
 \frac{dN}{dt\ dA\ dE_\nu\ d\Omega}=B f_{tot}[E_\nu]\ \frac{E_\nu^2} {c^2 \hbar^3 \pi^2},
\label{nf}
\ee\
where $B$ is a normalization constant set by
\be
\int_{E_\circ}^{E_P}  \frac{E_\nu\ dN}{dt\ dA\ dE_\nu\ d\Omega}\ dE_\nu= F^{tot}_{\nu}[z_{\rm max}].
\ee\

 We make the simplifying assumption that the neutrinos emitted from the set of stellar mass and supermassive black holes each have approximately the same minimum energy $E_\circ$ and thus the same spectrum.

An example of a power law with $p\sim2$ is first order Fermi acceleration \citep{fermi}. First order Fermi acceleration is a mechanism that describes the acceleration of charged particles, such as electrons, crossing strong shocks. Upon meeting the shock, due to the disturbance in the magnetic field caused by the shock, there is a probability for the incident electrons to get bumped back the way they travelled. There is also a probability for them to pass through the shock. For those charged particles that bounce back, this process occurs repeatedly until they escape. Each bounce causes the particle to gain energy and the statistics of this process yields a spectrum of:

\be
\frac{dN}{dE_\nu}=\frac{N_\circ} {E_\circ}\ \left(\frac{E_\circ}{E_\nu}\right)^{p}
\label{fa}
\ee
where $p\simeq 2$ in the nonrelativistic case \citep[in the relativistic case, the spectrum is still a power law but the index can vary depending on the details of the scattering; see e.g.,][]{vietri}. $N_\circ$ and $E_\circ$ are the initial number and energies of the charged particles. 

%

Fermi acceleration is ubiquitous for charged particles in astrophysical environments, due to their tight coupling to the magnetic field. Even though neutrinos are not charged and do not interact with an electromagnetic field, they do interact with a gravitational field which varies significantly on short time/length scales close to the firewall. Therefore, depending on the radiative properties of the firewall, it is possible for neutrinos to bounce back and forth between the accretion flow and the black hole firewall (through gravitational scattering), which could yield a power law whose precise index would depend on the details of the propagation model. We shall next discuss a possible mechanism for {\it gravitational Fermi acceleration}.

\subsection{Gravitational Fermi Acceleration}

Fermi acceleration requires presence of converging magnetic {\it mirrors}, which can generically occur in  a magnetohydrodynamic (MHD) turbulent medium (2nd order), or around strong shock fronts (1st order).  For neutral particles such as neutrinos, a similar effect can happen due to gravitational interactions.  In particular, relativistic neutrinos can only be significantly deflected by the gravitational field in the vicinity of black holes. However, null geodesics around ordinary Kerr black holes, generically either fall into the horizon or escape to infinity, and thus don't have the opportunity to up-scatter to higher energies. 

The picture can be different if we replace the black hole horizon by a firewall. In particular, one possibility that gives a local description of Bekenstein-Hawking entropy for firewalls excises the spacetime inside the stretched horizon, imposing $Z_2$ boundary conditions \citep{mehdi}. This implies that the firewall may act as a {\it mirror} for test particles. As motivated in Sec. \ref{intro}, we shall assume that this picture is indeed valid for neutrinos due to their weak interaction with matter. However, photons can be absorbed by the firewall, due to their electromagnetic interactions. 

We now have our two gravitational mirrors, one provided by the gravity of the black hole, and the other by a(n exotic) firewall. We shall next study the evolution of neutrino energy due to the fluctuations of spacetime geometry, caused by e.g. matter accreting onto a black hole. The energy is given by:
\be
E = \xi^\mu p_\mu,
\ee
 where $\xi^\mu \equiv \delta^\mu_0$ would be the time-like Killing vector for a stationary metric, while $p_\mu$ is the four-momentum.  For a time-dependent metric, the rate of energy change is given by:
 \be
 dE = \frac{1}{2} dx^\mu p^\nu \xi_{(\mu;\nu)} = - \frac{1}{2} \left( g_{\mu\nu,0}dx^\mu dx^\nu \over g_{\alpha\beta}\xi^\alpha dx^\beta \right) E,
 \ee
 i.e. the relative change in particle energy is roughly proportional to the relative change in the metric:
 \be
 \Delta \ln E = -\frac{1}{2} \int d\lambda \left( g_{\mu\nu,0}u^\mu u^\nu \over g_{\alpha\beta}\xi^\alpha u^\beta \right) = {\cal O} (\Delta \ln |g_{\mu\nu}|),
 \ee
 where $u^\mu = dx^\mu/d\lambda$, is defined using an affine parameter $\lambda$. 
 
 The change in particle energy can be divided up into a systematic drift, and a random walk. The systematic change can be calculated using {\it adiabatic invariance}, i.e. that action variables for bound orbits of integrable systems (away from the resonances):
 \be
 J_i  \equiv \oint p_i dx^i, 
 \ee
 remain invariant under slow changes of the Hamiltonian.  Geodesics in Kerr spacetime are indeed integrable, as three integrals of motion \citep[energy, azimuthal angular momentum, and Carter constant;][]{1968PhRv..174.1559C} exist.  For null geodesics, the action variables simply scale as $J \sim E\times R \propto E \times M_{\rm BH}$, so adiabatic invariance yields:
 \be
 \langle \Delta \ln E \rangle \simeq - \Delta \ln M_{\rm BH}, \label{drift}
 \ee 
 i.e. bound relativistic particles systematically redshift with the size of the black hole horizon. 
 
 The computation of the random contribution to gravitational redshift is more involved, as it depends on the details of metric fluctuations induced by the turbulent accretion disc. However, this could simply be measured by studying geodesics of test particles in existing general relativistic magnetohydrodynamic  (GRMHD) simulations (with live geometry) of accretion discs in their infalling regions. Let's try to estimate this:
 
Given that an accretion disc of radius $R$  is turbulent on the scale of its scale-height, $H$,  we estimate the relative change in energy of a particle in one orbit to be:
\be
\Delta \ln E \sim  \Delta \ln |g_{\mu\nu}| \sim \pm \frac{H}{R} \times \frac{m_{\rm disc}}{M_{\rm BH}}. 
\ee
Over several orbits, $\ln E$ would undergo a random walk (plus the systematic drift in \eqref{drift}), with the variance:
\be
 \langle \left(\Delta \ln E\right)^2 \rangle \sim \left(H \over R\right)^2 \left(\frac{m_{\rm disc}}{M_{\rm BH}}\right)^2 \frac{t}{t_{\rm orbit}}.
\ee
On the other hand, the accretion rate (for a radiatively efficient accretion disc) is roughly given by:
\be
\dot{M}_{\rm BH} \sim \alpha_{SS} m_{\rm disc} \left(2 \pi \over t_{\rm orbit}\right) \left(H \over R\right)^2, \label{acc_rate}
\ee
where $\alpha_{\rm SS} \sim 0.01-0.1$ is the celebrated Shakura-Sunyaev viscosity parameter \citep[e.g., see][]{Balbus}. This yields:
\be
 \langle \left(\Delta \ln E\right)^2 \rangle \sim  \frac{1}{2\pi \alpha_{SS}} \times \frac{m_{\rm disc}}{M_{\rm BH}} \times \Delta\ln M_{\rm BH}.\label{r_walk}
 \ee
 
 Equipped with Eqs. (\ref{drift}) and (\ref{r_walk}), we can write a Fokker-Planck equation for the evolution of the energy distribution of neutrinos, trapped between a firewall and the gravitational potential barrier of an accreting black hole:
 \be
\frac{ \partial n}{\partial t} = \frac{\partial}{\partial \ln E_\nu } \left[   -A n +  \frac{1}{2} B \frac{\partial n}{\partial \ln E_\nu }  \right], \label{FP}
\ee
where 
\bea
n &\equiv& E_\nu \frac{dN}{dE_\nu} = \frac{dN}{d\ln E_\nu}, \\
A &\equiv& \frac{d}{dt} \langle \Delta \ln E \rangle = -\frac{\dot{M}_{\rm BH}}{M_{\rm BH}}, \\
B &\equiv&  \frac{d}{dt} \langle (\Delta \ln E)^2 \rangle =  \frac{1}{2\pi \alpha_{SS}} \times \frac{m_{\rm disc}}{M_{\rm BH}} \times \frac{\dot{M}_{\rm BH}}{M_{\rm BH}}.\nonumber\\ &&
\eea

The equilibrium solution to the Fokker-Planck equation (\ref{FP}), assuming zero flux through (high or low energy) boundaries, gives:
\be
p = 1- \frac{\partial \ln n}{\partial \ln E} =  1- \frac{2A}{B} \sim 1 + \frac{4\pi \alpha_{SS}}{m_{\rm disc}/M_{\rm BH}}, 
\ee
or equivalently
\be
p-1 \sim 1.3 \left( \alpha_{\rm SS} \over 0.01 \right) \left(  m_{\rm disc}/M_{\rm BH} \over 0.1 \right)^{-1}. 
\ee

The requirement of a significant disc to black hole mass fraction ($\gtrsim 10\%$), to fit the IceCube measurement of $p \sim 2.5$, suggests a stellar origin, e.g., the fallback discs of gamma ray bursts. Alternatively, the accretion rate (\ref{acc_rate}) could be significantly smaller for radiatively {\it inefficient} flows \citep[e.g.,][]{Yuan:2003dc}, but the details will be more model-dependent in this regime.

Note that this calculation provides the steady-state spectrum of trapped neutrinos. However, we expect the spectrum of neutrinos that escape to infinity to be the same, as the escape probability of relativistic particles is independent of energy, and only depends on geometric factors.

Alternatively, protons could be accelerated via first order Fermi mechanism, and then produce neutrinos through hadronic processes (e.g. \citet{Waxman:2001kt}). We will consider indices $p\sim 2-3$. \\

In general, we think that a power law with $p\sim2$ is a natural choice. This is because high energy processes (such as possible Planck-scale physics within firewalls) favour hard spectra, but the requirement of convergence sets the lower  limit of $p= 2$ (equal energy per decade).
%
%
%

\begin{figure*}
\includegraphics[width=1.09\textwidth]{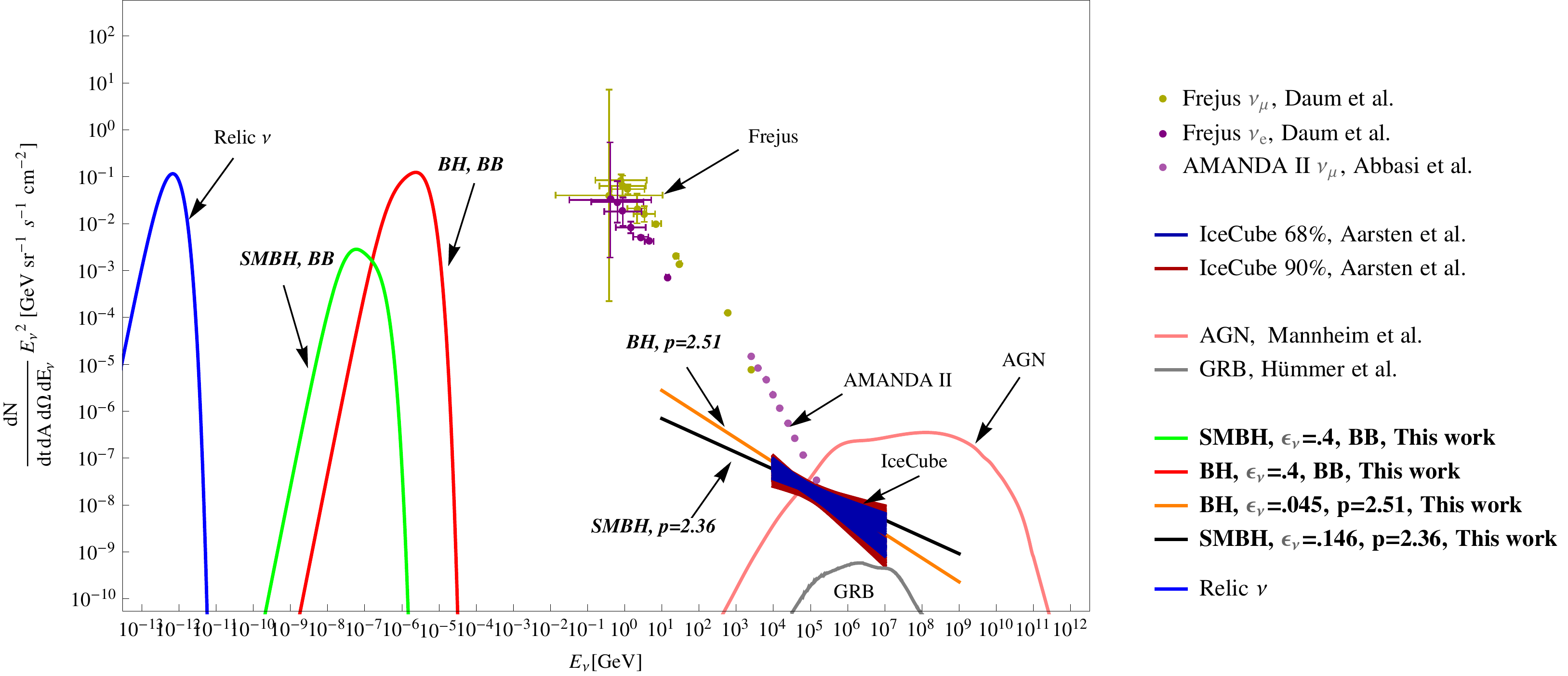}
\caption{Models and observations of neutrino fluxes. The AGN and gamma ray burst (GRB) predictions are from \citet{agn} and \citet{grb} respectively. The IceCube detection fits are from \cite{icecube1}, and the atmospheric neutrino detections are from  Fr\'{e}jus \citep{frejus} and AMANDA-II \citep{amandaii}. The predictions for the neutrino fluxes from surfaces of black holes with a blackbody and power law spectrum are from this work.}
\label{flx}
\end{figure*}

\begin{figure*}
\includegraphics[width=.9\textwidth]{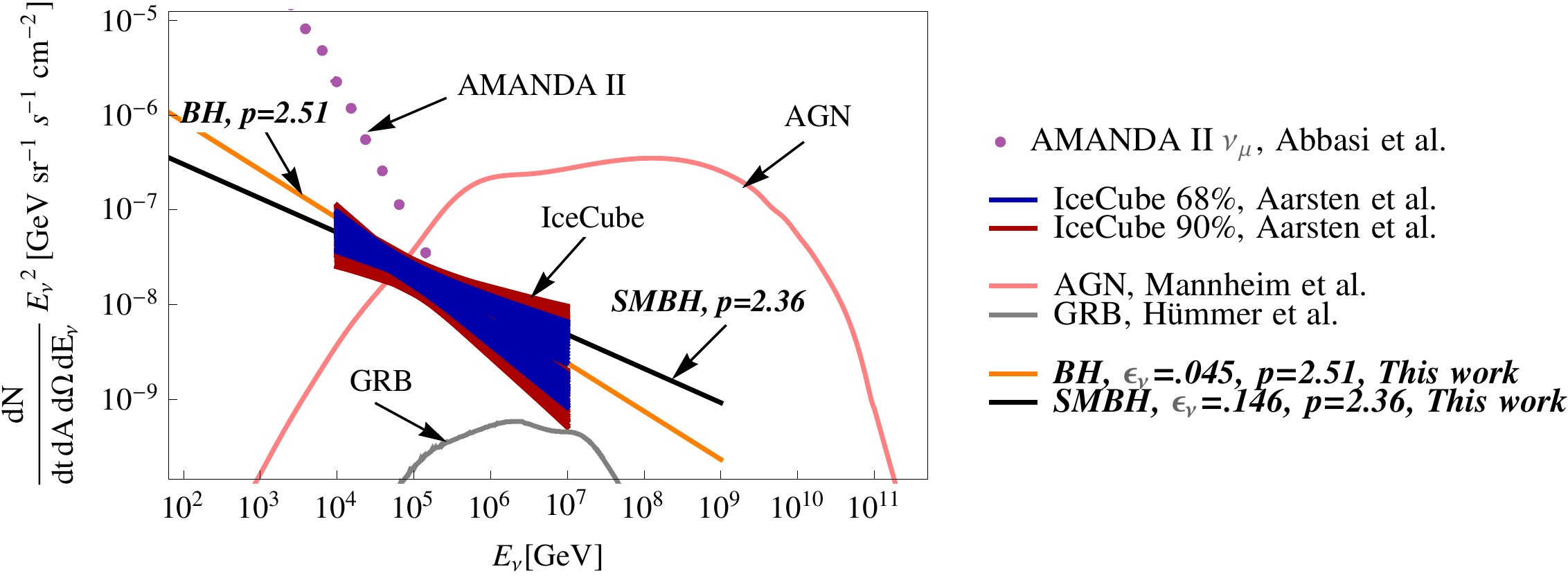}
\caption{A close-up of Fig. (\ref{flx}) within the vicinity of the IceCube detection. }
\label{zoomflx}
\end{figure*}

\subsection{Blackbody}\label{sec:bb}
If we assume that the black holes are in approximate steady state and are close to thermodynamic equilibrium, we can consider them as blackbodies. Again, we make the simplifying assumptions that the set of stellar mass and supermassive black holes each have approximately the same surface temperature $T$ and thus spectrum. The temperature can be determined by the neutrino Stefan-Boltzmann law:

\be
L_{\nu, BB} \sim 4\pi\sigma r_A^2\ T^4
\ee

 With $\epsilon_\nu=0.4$, we find $T_{\rm BH}=1.2\times 10^7$ K and $T_{\rm SMBH}=6.7\times 10^5$ K. The energy distribution is:
\be
\frac{dN}{dE_\nu}=\frac{1}{\pi^2 c^3\hbar^3}\frac{E_\nu^2}{e^{E_\nu/kT}+1}.
\ee

The number flux is again obtained using \eqref{nf}, with the phase space density
\be
f_{tot}[E_\nu]=\int^{z_{\rm max}} \left(\frac{\dot{\rho}(z) c^2 dV }{L_{\nu, BB}}\right) \left(\frac{\pi r_A^2}{4\pi d_A^2}\right) \frac{1}{1+e^{E_\nu(1+z)/kT}}.
\ee


 \begin{figure}
      \centering
           \includegraphics[width=.8\linewidth]{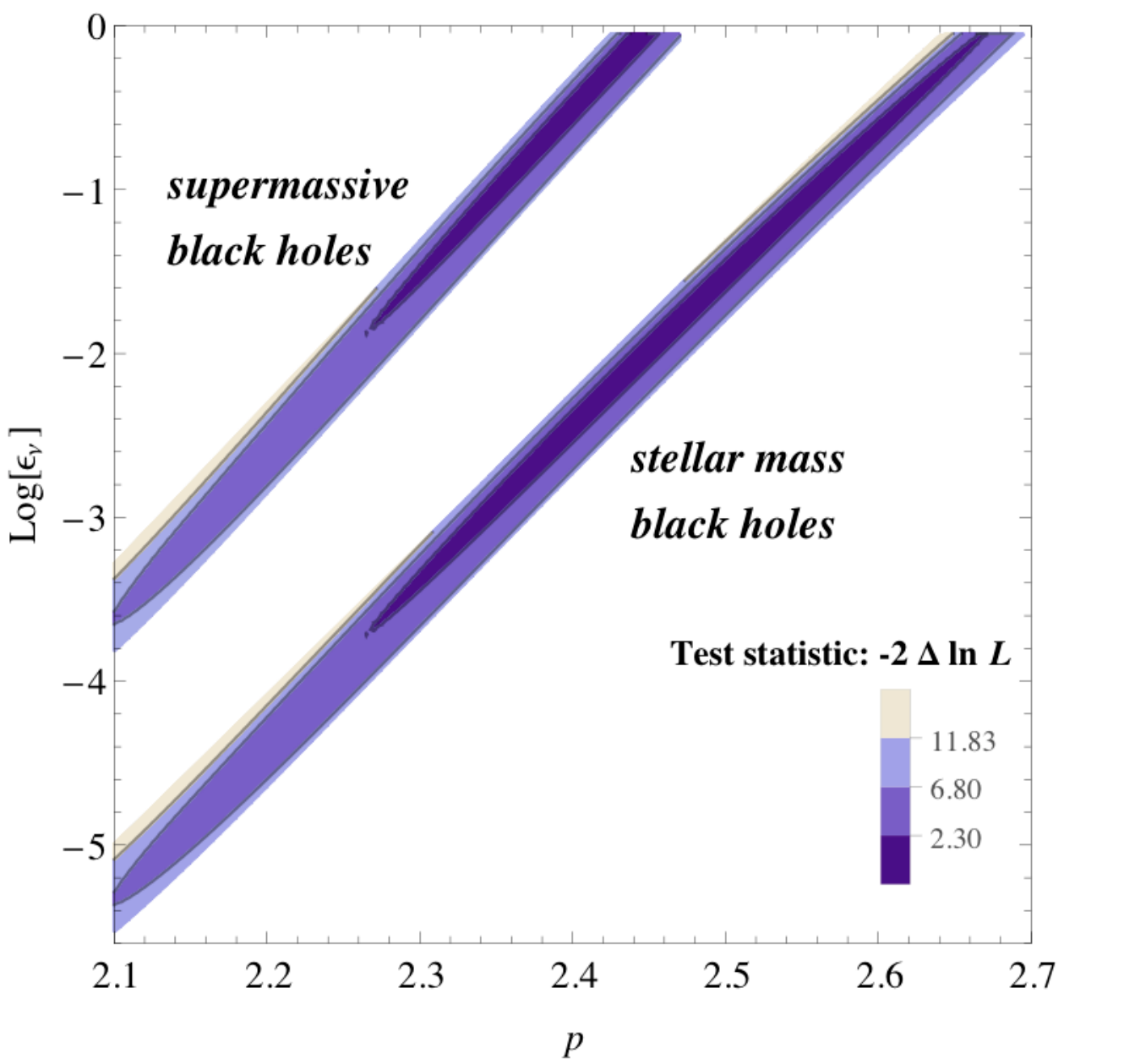}

 \caption{IceCube likelihood contours for firewall neutrino emission efficiency $\epsilon_\nu$ and spectral index $p$, for supermassive {\it(Upper)} and stellar mass {\it(Lower)} black holes. The shaded regions from dark to light correspond to the $1\sigma$, $2\sigma$, and $3\sigma$ contours. These contours are obtained from a likelihood-ratio test using the test statistic $-2\Delta\ln L=-2(\ln L-\ln L_{max})$.}
\label{bhlike}
    \end{figure}

\section{Results}\label{sec:results}

The IceCube collaboration has recently observed a flux of PeV ($10^{15}$ eV) neutrinos \citep{icecube1}, and the source of these high energy neutrinos is as yet unknown. There are only upper bounds beyond ~2 PeV, and these are consistent with the power laws we consider in this work.

Figure \ref{flx} shows our results for the neutrino spectra for the stellar mass black holes and the supermassive black holes, for the two cases of power law (with $p\sim2$) and blackbody spectra, as discussed in the previous section (Note that, Figure \ref{flx} shows only a sample pair of $\epsilon_\nu, p$ values that fit the data. For the complete set of parameters that fit the IceCube data, see Figure \ref{bhlike}). Also included in the figure are neutrino flux predictions from models of AGN \citep{agn} and GRB's \citep{grb}, along with a fit to IceCube's detection from \citep{icecube1}, and atmospheric neutrino detections. The atmospheric neutrino observations are from Fr\'{e}jus \citep{frejus} and AMANDA-II \citep{amandaii}. The blackbody models do not have interesting contributions to the high energy spectra, but it would be interesting in future work to compare these results to upper limits on other sources of low energy neutrinos. In this work, we are interested in high energy neutrinos, so we focus on the power law model results. In Figure \ref{zoomflx}, we have zoomed in on our results and the other predictions, in the vicinity of the IceCube detection. We see the firewall neutrino signal can certainly fit the IceCube detection and is a possible candidate for the source of these high energy neutrinos.

Figure \ref{bhlike}  shows the parameter combinations for neutrino emission efficiency $\epsilon_\nu$ and spectral index $p$ for which our power law model yields a good fit to the IceCube data. For the power law models, fits to the IceCube's data are within the physically acceptable range ($0<\epsilon_g\leq 1$) at $96\%$ and $47\%$ probability for stellar mass black holes and supermassive black holes respectively (or $92\%$ and $37\%$ for $0<\epsilon_g\leq 0.5$). These probabilities can change by $\pm 5\%$ across different models for star formation  and supermassive black hole growth. At the $1\sigma$ confidence level the fit parameters for stellar mass (supermassive) black holes are $\epsilon_\nu\gtrsim 10^{-4} ~ (10^{-2})$ and  $p=2.26-2.68 ~ (2.27-2.45)$, while at the $2\sigma$ level they are $\epsilon_\nu\gtrsim 10^{-5} ~ (10^{-3.5})$ and $p=2.1-2.7 ~ (2.1-2.47)$.

\section{Conclusions}\label{sec:conclusions}

To summarize, we have argued that accretion onto astrophysical firewalls is likely to predominantly lead to a flux of cosmological neutrinos. If this spectrum has a power-law shape, e.g. due to an analog of Fermi acceleration for neutrinos, then it could well fit the observed IceCube spectrum of high energy neutrinos at reasonable efficiencies from astrophysical black holes. Future work should focus on more detailed spectral modelling of neutrino emission from putative firewall atmospheres, as well as other possible observational smoking guns for this scenario. It is certainly exciting to entertain the possibility that neutrino astrophysics could open a new window into the physics of quantum gravity and black holes. \\

\section*{Acknowledgments}

We thank Julia Becker Tjus for giving us the AGN model data that we compare our results to in this paper, and we thank Jakob van Santen for providing us with IceCube's likelihood profiles. We also thank Avery Broderick and Ramesh Narayan for fruitful discussions, as well as John Beacom, Mauricio Bustamante, Avi Loeb and Ue-Li Pen for comments on the manuscript. This research was supported in part by Perimeter Institute for Theoretical Physics (PI). Research at PI is supported by the Government of Canada through Industry Canada and by the Province of Ontario through the Ministry of Research and Innovation.


\begin{thebibliography}{99}
    \bibitem[\protect\citeauthoryear{Aartsen et al.}{2014}]{icecube3}
Aartsen E.{\sl et al.} (IceCube Collaboration), Phys. Rev. Lett. 113, 101101 (2014).

    \bibitem[\protect\citeauthoryear{Aartsen et al.}{2014}]{icecube1}
    Aartsen E. {\sl et al.} (IceCube Collaboration), arXiv:1410.1749, (2014).
    \bibitem[\protect\citeauthoryear{Abbasi et al.}{2010}]{amandaii}
 Abbasi R.{\sl et al.}  (IceCube Collaboration), Astropart. Phys. 34, 48 (2010).

\bibitem[\protect\citeauthoryear{Almheiri et al.}{2013}]{firewall1}
    Almheiri A., Marolf D., Polchinski J., Sully J., Journal of High Energy Physics, 2013, 2, (2013).
\bibitem[\protect\citeauthoryear{Avery, Chowdhury \& Puhm}{2013}]{y22} Avery S. G., Chowdhury  B. D., Puhm A., arXiv:1210.6996 [hep-th].
\bibitem[\protect\citeauthoryear{Balbus \& Hawley}{1998}]{Balbus} 
  Balbus S. A., Hawley J. F.,
  Rev.\ Mod.\ Phys.\  {\bf 70}, 1 (1998).
\bibitem[\protect\citeauthoryear{Banks \& Fischler}{2012}]{y29}Banks T., Fischler W., arXiv:1208.4757 [hep-th].    
 \bibitem[\protect\citeauthoryear{Banks \& Fischler}{2013}]{y7} Banks T., Fischler W., arXiv:1305.3923 [hep-th].
\bibitem[\protect\citeauthoryear{Bena, Puhm \& Vercnocke}{2012}]{y27}Bena I., Puhm A., Vercnocke B., JHEP 1212 , 
014 (2012) [arXiv:1208.3468 [hep-th]].

\bibitem[\protect\citeauthoryear{Bousso}{2012}]{y34} Bousso R., arXiv:1207.5192 [hep-th].
 \bibitem[\protect\citeauthoryear{Braunstein, Pirandola \& Zyczkowski}{2013}]{y2} Braunstein S., Pirandola S., Zyczkowski K., Physical Review Letters. 2013 Mar 8;110(10). 101301.
       \bibitem[\protect\citeauthoryear{Broderick, Loeb \& Narayan}{2009}]{avery}
       Broderick A. E., Loeb A., Narayan R., ApJ, 701, 1357, (2009).
       
         \bibitem[\protect\citeauthoryear{Broderick et al.}{2015}]{2015ApJ...805..179B} 
  Broderick A.~E., Narayan R., Kormendy J., Perlman E.~S., Rieke M.~J., Doeleman S.~S., 2015, ApJ, 805, p 


\bibitem[\protect\citeauthoryear{Brustein}{2012}]{y25}Brustein R., arXiv:1209.2686 [hep-th].
\bibitem[\protect\citeauthoryear{Carter}{1968}]{1968PhRv..174.1559C} 
  Carter B.,  Phys. Rev. , 174 (1968).    

\bibitem[\protect\citeauthoryear{Chowdhury \& Puhm}{2012}]{y31} Chowdhury B. D., Puhm A., arXiv:1208.2026 [hep-th].
    \bibitem[\protect\citeauthoryear{Daum et al.}{1995}]{frejus}
    Daum K.{\sl et al.}  Zeitschrift f\H{u}r Physik C, 66:417, (1995).

  \bibitem[\protect\citeauthoryear{Di Matteo et al.}{2003}]{SH1}
  Di Matteo T., Croft R. A. C., Springel V., Hernquist L., The Astrophysical Journal, 593, 1, (2003).

\bibitem[\protect\citeauthoryear{Dodelson \& Silverstein}{2015}]{2015arXiv150405536D} 
  Dodelson M., Silverstein E., 2015, arXiv:1504.05536 

       \bibitem[\protect\citeauthoryear{Fermi}{1949}]{fermi}
    Fermi E., Physical Review 75, (1949). 	arXiv:1410.1749

\bibitem[\protect\citeauthoryear{Frank, King \& Raine}{2002}]{king} 
  Frank J., King A., Raine D.~J., \emph{Accretion Power in Astrophysics: Third Edition}, Cambridge University Press, (2002). 

\bibitem[\protect\citeauthoryear{Fukugita \& Peebles}{2004}]{cei}
Fukugita M., Peebles P. J. E., Astrophys.J.616:643-668, (2004).

\bibitem[\protect\citeauthoryear{Giddings}{2013}]{y17} Giddings S. B., arXiv:1211.7070 [hep-th].
\bibitem[\protect\citeauthoryear{Giddings}{2013}]{y10} Giddings S. B. , arXiv:1302.2613 [hep-th].
 \bibitem[\protect\citeauthoryear{Giddings}{2014}]{2014PhRvD..90l4033G} Giddings S.~B., 2014, PhRvD, 90, 124033.
\bibitem[\protect\citeauthoryear{Giveon \& Itzhaki}{2012}]{y26}Giveon A., Itzhaki N., JHEP 1212, 094 (2012) [arXiv:1208.3930 [hep-th]].
\bibitem[\protect\citeauthoryear{Harlow \& Hayden}{2013}]{y13} Harlow D., Hayden P., JHEP1306, 085 (2013) [arXiv:1301.4504 [hep-th]].

\bibitem[\protect\citeauthoryear{Hossenfelder}{2012}]{y23} Hossenfelder S., arXiv:1210.5317 [gr-qc].
\bibitem[\protect\citeauthoryear{Hsu}{2013}]{y11}Hsu S. D. H., arXiv:1302.0451 [hep-th].
    \bibitem[\protect\citeauthoryear{H\"{u}mmer, Baerwald \& Winter}{2012}]{grb}
H\"{u}mmer S.,  Baerwald P., Winter W., Phys. Rev. Lett. 108, 231101 (2012).

 \bibitem[\protect\citeauthoryear{Hutchinson \& Stojkovic}{2013}]{y3}  Hutchinson J., Stojkovic D., arXiv:1307.5861.
\bibitem[\protect\citeauthoryear{Hwang, Lee \& Yeom}{2013}]{y20} Hwang  D. I., Lee  B. H., Yeom D.H., JCAP 1301, 005 (2013)[arXiv:1210.6733 [gr-qc]].
\bibitem[\protect\citeauthoryear{Jacobson}{2014}]{y15} Jacobson T. , arXiv:1212.6944 [hep-th].
\bibitem[\protect\citeauthoryear{Kawai, Matsuo \& Yokokura}{2013}]{y9} Kawai H., Matsuo Y., Yokokura Y., Int. J. Mod. Phys. A28, 1350050 (2013)[arXiv:1302.4733 [hep-th]].
\bibitem[\protect\citeauthoryear{Kim, Lee \& Yeom}{2013}]{y12}Kim W., Lee B. H., Yeom D. H., JHEP1305, 060 (2013) [arXiv:1301.5138[gr-qc]].

\bibitem[\protect\citeauthoryear{Larjo, Lowe \& Thorlacius}{2013}]{y16} Larjo K., Lowe  D. A., Thorlacius L., Phys. Rev. D 87, 104018(2013)[arXiv:1211.4620 [hep-th]].
\bibitem[\protect\citeauthoryear{Lee \& Yeom}{2013}]{y8} Lee B. H, Yeom D. H., arXiv:1302.6006 [gr-qc].
 \bibitem[\protect\citeauthoryear{Lowe \& Thorlacius}{2013}]{y6} Lowe D. A., Thorlacius L., arXiv:1305.7459 [hep-th].
  \bibitem[\protect\citeauthoryear{Mannheim, Protheroe \& Rachen}{2001}]{agn}
Mannheim K., Protheroe R. J., Rachen J. P., Phys. Rev. D 63, 023003 (2001).
  
\bibitem[\protect\citeauthoryear{Mathur \& Turton}{2014}]{y32} Mathur S. D., Turton D., arXiv:1208.2005 [hep-th].
\bibitem[\protect\citeauthoryear{Murase, Ahlers \& Lacki}{2013}]{murase}
Murase K., Ahlers M., Lacki B. C. , PRD 88, 121301 (2013).
 
 \bibitem[\protect\citeauthoryear{Murase et al.}{2014}]{murase2} K. Murase, Y. Inoue, C. D. Dermer, Phys.Rev. D90 023007, (2014).

\bibitem[\protect\citeauthoryear{Nomura \& Varela}{2013}]{y18} Nomura Y., Varela J., arXiv:1211.7033 [hep-th].
\bibitem[\protect\citeauthoryear{Nomura, Varela \&Weinberg}{2013}]{y33} Nomura Y., Varela J., Weinberg S. J., JHEP 1303, 059 (2013)[arXiv:1207.6626 [hep-th]].

\bibitem[\protect\citeauthoryear{Nomura, Varela \& Weinberg}{2013}]{y21} Nomura Y.,Varela J., Weinberg S. J., Phys. Rev. D 87, 084050 (2013) [arXiv:1210.6348 [hep-th]].
\bibitem[\protect\citeauthoryear{Ori}{2012}]{y28} Ori A., arXiv:1208.6480 [gr-qc].
 \bibitem[\protect\citeauthoryear{Page}{2013}]{y5}Page D. N., arXiv:1306.0562 [hep-th].
 \bibitem[\protect\citeauthoryear{Pen \& Broderick}{2014}]{y4} Pen U.-L., Broderick A.~E., 2014, MNRAS, 445. 

\bibitem[\protect\citeauthoryear{Rama}{2014}]{y19} Rama S. K., arXiv:1211.5645 [hep-th].
 \bibitem[\protect\citeauthoryear{Saravani, Afshordi \& Mann}{2015}]{mehdi}
 Saravani M., Afshordi N., Mann R. B.,
  Int.\ J.\ Mod.\ Phys.\ D {\bf 23}, no. 13, 1443007 (2015)
  [arXiv:1212.4176 [hep-th]].
 \bibitem[\protect\citeauthoryear{Springel \& Hernquist}{2003}]{SH2}
 Springel V., Hernquist L., Monthly Notice of the Royal Astronomical Society, 339, 2, (2003).

\bibitem[\protect\citeauthoryear{Susskind}{2012}]{y30}Susskind L., arXiv:1208.3445 [hep-th].
\bibitem[\protect\citeauthoryear{Susskind}{2012}]{y24}Susskind L., arXiv:1210.2098 [hep-th].
\bibitem[\protect\citeauthoryear{Susskind}{2013}]{y14} Susskind L., arXiv:1301.4505 [hep-th].

\bibitem[\protect\citeauthoryear{Vaz}{2015}]{y1} Vaz C., arXiv:1407.3823 [gr-qc].
    \bibitem[\protect\citeauthoryear{Vietri}{2008}]{vietri}
    Vietri M., \emph{Foundations of High-Energy Astrophysics}, Univ. Chicago Press, (2008).

    \bibitem[\protect\citeauthoryear{Waxman \& Loeb}{2011}]{Waxman:2001kt} 
  Waxman E., Loeb A.,
  Phys.\ Rev.\ Lett.\  {\bf 87}, 071101 (2001)
  [astro-ph/0102317].

  \bibitem[\protect\citeauthoryear{Yoshida \& Takami}{2014}] {yoshida}
    Yoshida S., Takami H., arXiv:1409.2950, (2014).  

  
  \bibitem[\protect\citeauthoryear{Yuan, Quataert \& Narayan}{2003}]{Yuan:2003dc} 
Yuan F., Quataert E.,Narayan R.,
  Astrophys.\ J.\  {\bf 598}, 301 (2003)
  [astro-ph/0304125].












\end{thebibliography}
\end{document}